\documentclass[a4paper,twoside]{article}
\usepackage{here}
\usepackage{epsfig}
\usepackage{color}
\usepackage{subcaption}
\usepackage{calc}
\usepackage{amssymb}
\usepackage{amstext}
\usepackage{amsmath}
\usepackage{amsthm}
\usepackage{multicol}
\usepackage{pslatex}
\usepackage{apalike}
\usepackage{SCITEPRESS}

\begin{document}

\title{Cell Image Segmentation by Feature Random Enhancement Module}

\author{\authorname{Takamasa Ando \sup{1}, Kazuhiro Hotta\sup{2}}
\affiliation{\sup{1,2}Meijo University,1-501 Shiogamaguchi, Tempaku-ku, Nagoya 468-8502, Japan}
\email{\sup{1}203427003@ccalumni.meijo-u.ac.jp, \sup{2}kazuhottta@meio-u.ac.jp}
}


\abstract{It is important to extract good features using an encoder to realize semantic segmentation with high accuracy. Although loss function is optimized in training deep neural network, far layers from the layers for computing loss function are difficult to train. Skip connection is effective for this problem but there are still far layers from the loss function. In this paper, we propose the Feature Random Enhancement Module which enhances the features randomly in only training. By emphasizing the features at far layers from loss function, we can train those layers well and the accuracy was improved. In experiments, we evaluated the proposed module on two kinds of cell image datasets, and our module improved the segmentation accuracy without increasing computational cost in test phase. 
}

\onecolumn \maketitle \normalsize 

\begin{figure*}[t]
\centering
\includegraphics[height=4.5cm]{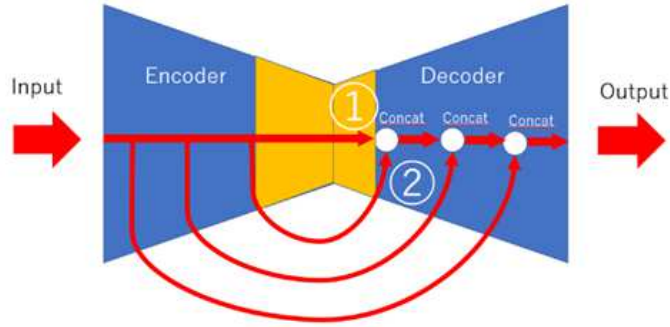}
\caption{U-Net and the problem}
\end{figure*}

\section{\uppercase{Introduction}}
\label{sec:introduction}

In recent years, the development of deep learning technology has been remarkable, and there is a demand to use it in various situations. Since the segmentation of cell images obtained by microscopes is performed manually by human experts, it tends to be subjective results.
Objective results are required by the same criteria using deep learning technology \cite{one}. However, 
the optimal network for segmentation using deep learning has not been established yet. Even if the accuracy is not so high, it is actually used in the field of cell biology to obtain objective results. Therefore, automatic segmentation method with high accuracy is desired. U-Net is still widely used for segmentation of microscope images because it works well for small number of training images and high accuracy is obtained without adjusting hyperparameters. For this reason, many improvements of U-Net have been proposed for microscope images \cite{six,seven,eight}. 

This study belongs to one of those variations and improves the accuracy of U-Net. Although conventional improvement is done by deepening, the proposed method does not require any additional computational resources at all during inference. Therefore, it retains the advantage of U-Net that it requires fewer computing resources. Therefore, it is a very significant proposal in the segmentation of cell images where there is a demand for lightweight and accurate model.

A neural network such as CNN basically uses backpropagation for training. For this reason, there is a phenomenon that near layers to the layer for computing loss are more updated in comparison with far layers from the loss \cite{three}. In order to solve the problem, ResNet \cite{four} used skip connection and improved the accuracy. U-Net \cite{two} is the famous deep neural network for cell segmentation task. U-Net also has skip connection between encoder and decoder, and it contributes to improve the accuracy. In general, it is well known that skip connection gives the information of location and fine objects which were lost in encoder to decoder. However, we consider that the same theory as ResNet is used in skip connection to improve the accuracy. By using skip connection, the loss is propagated to encoder well, and weights are successfully updated. This is also the reason why U-net improved the accuracy in comparison with standard Encoder-Decoder CNN. 

Figure 1 shows the structure of U-net. We see that skip connection is effective to propagate the loss to encoder. However, the layers shown as yellow in the Figure 1 are the farthest from the loss at the final layer. Therefore, in the case of U-net, the yellow layer in the Figure is the most difficult to train though the layer has semantic information. In this paper, we propose new module to train the layer effectively. 
We consider to enhance the feature map at yellow layer which is the farthest layer from the loss function. In conventional methods, the yellow layer is difficult to learn, and network learns to decrease the feature values at the yellow layer not to affect the output. This phenomenon is shown in section 3.1.
Feature values at the yellow layer are smaller than those at skip connection from encoder, and features at the yellow layer are not effective for segmentation result. 
Therefore, the model without yellow layer sometimes has higher accuracy than the model with yellow layer as shown in section 4.3. 
To train yellow layers effectively, we select some feature maps randomly at yellow layer and increase the absolute value of the feature maps multiplied by a large constant value. Although the features at yellow layer and features from encoder are concatenated, the features at yellow layer are used mainly because the yellow layer has larger values by constant multiplication. 
If the filters selected by our module are not effective for classification, network has a large loss. Therefore, the selected feature maps are trained well by gradient descent.

In experiments, we evaluated our method on two kinds of cell image datasets. Intersection over Union (IoU) is used as an evaluation measure. The effectiveness of the proposed module was shown in comparison with the conventional U-Net without our module and U-net with SuperVision that loss function is computed at yellow layer. 

The structure of this paper is as follows. Section 2 describes related works. Section 3 describes the details of the proposed method. Experimental result on two kinds of cell image datasets are shown in section 4. Finally, we summarize our work and discusses future works in section 5.

\begin{figure*}[t]
\centering
\includegraphics[height=4.0cm]{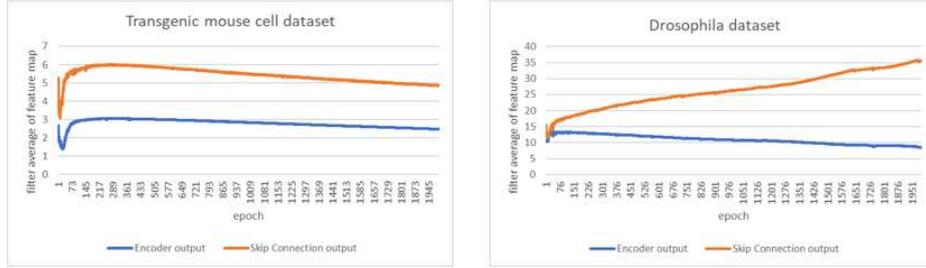}
\caption{Comparison of feature values at the yellow layer and skip connection}
\end{figure*}

\section{\uppercase{Related works}}

U-Net is a kind of Encoder-Decoder CNN \cite{five}. Unlike the PSPNet \cite{nine}, the Encoder-Decoder CNN does not use features in parallel but features are extracted in series. Thus, in Encoder-Decoder CNN, far layers from the layer for computing loss are not updated well. ResNet and U-net solved this problem by skip connection.

There is also a technique called Deep supervision proposed in Deeply-Supervised Nets \cite{ten} to address the problem. In deep supervision, loss is also computed at middle layer. Far layers from final layer are updated well by supervision. U-Net++ \cite{eleven} also used this technique. However, forcing loss from the ground truth in the middle of U-Net may not obtain an intermediate representation for better inference. In addition, U-Net++ has a structure in which the output image is restored by the decoder from various parts of the encoder, and multiple decoders are connected to each other. However, the advantage of U-Net which is a small computational resource is vanished. This is accompanied by a large number of parameters due to multiple decoders. In this paper, we propose new methods based on the merits and demerits of these previous studies.

There are some methods that we referred to consider a new method. In the proposed method, feature enhancement is performed on some feature maps during training. There are many techniques for weighting feature maps. SENet \cite{twelve} proposed to weight important channels. Attention which has been proposed in the field of natural languages \cite{thirteen} is also used in the field of images. In recent years, many methods have been proposed that focus on channels \cite{fourteen,fifteen,sixteen}. Attention-U-net used attention for skip connection \cite{seventeen}. 

Dropout \cite{eighteen} is also related to our approach. Dropout sets a part of the feature map to 0 in only training. This prevents overfitting by randomly removing elements in only training. Our method randomly enhances some feature maps at the farthest layer from loss function. We do not set some elements to 0 and enlarge some feature maps. When some elements are set to 0,  backpropagation from the element is stopped. In our method, features are enlarged randomly to use backpropagation effectively for the farthest layer. 

\section{\uppercase{Proposed method}}

This section describes the proposed method. Section 3.1 gives the overview of the proposed method. Section 3.2 mentions the implementation details of our method. 

\subsection{Overview of the proposed method}

When we obtain segmentation result by the U-Net, the magnitude of features at yellow layer as shown in Fig. 1 is often smaller than that of features at skip connection from encoder to decoder. Fig.2 shows the fact when U-net is trained on two different datasets. 
Two lines in each graph show the average feature values at yellow layer in Fig.1 and those at skip connection from encoder to decoder. 
Note that both features are extracted after ReLU function and those two feature maps have positive values. 
The magnitude of these values in Fig.2 means the magnitude of influence for the network's outputs because convolution is adopted after both features in Fig.2 are concatenated.
The Figure shows that the encoder's output features are obviously smaller than the features at skip connection, and the features at yellow layer are not used effectively. 
Batch renormalization is used before we compute ReLU function in Fig 2. 
If we do not use backpropagation, both feature maps have similar average values because the features are normalized by batch renormalization.
Therefore, the training of U-net makes the features at yellow layer small.

Does this fact show that yellow layer is not required? Our answer is “NO”. This phenomenon is caused from that near layers to the layers for computing loss are updated well and far layers are not updated well. Yellow layer in Fig.1 is the farthest layer from loss function because encoder is updated through skip connection. Therefore, network learns the layers connected by a skip connection in comparison with the yellow layer because it is difficult to update the yellow layer.

Although the easy solution for this problem is to use new normalization for yellow layers using average and variance of skip connection, it prevents the yellow layer's values from being small. However, it generates new problem that can not learn appropriate ratio of yellow layer and skip connection values.
Therefore, we propose feature enhancement module to solve the problem. Our method is soft constraint as emphasis some feature maps in comparison with normalization. The soft constraint means emphasizing "a part of feature maps" at yellow layer. The new normalization described above increases the values in "all feature maps" at yellow layer to match the skip connection. Specifically, we select some feature maps at yellow layer randomly at each epoch and increase the absolute values of the features by multiplying a large constant value. This allows to use the features at yellow layer effectively. If the features enlarged by our module are not effective, network has a large loss. Therefore, the selected filters are trained well by gradient descent. The reason why we select some feature maps randomly is to prevent vanishing gradient.

Our method is not used in test phase. This is because it able to learn the appropriate magnitude of values in non-selected feature maps. The evidence for this is from experimental results, and it is shown in section 4.4. Since our method is soft constraint, it solve the problem that the normalization at the concatenation of two feature maps can not solve. Thus, the accuracy is improved without changing the inference time or computational resources because our method is not used in test phase. This is an advantage of our method though many methods deepened the network to improve the accuracy. 

\begin{figure*}
\centering
\includegraphics[height=5.0cm]{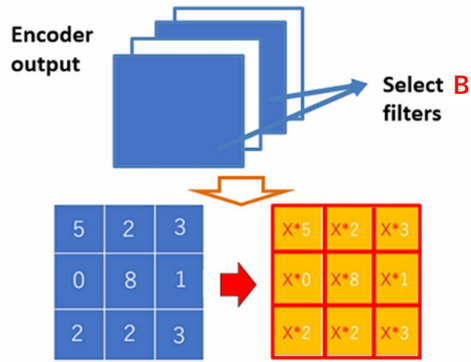}
\caption{Feature Random Enhancement Module}
\end{figure*}

\begin{figure*}
\centering
\includegraphics[height=5.0cm]{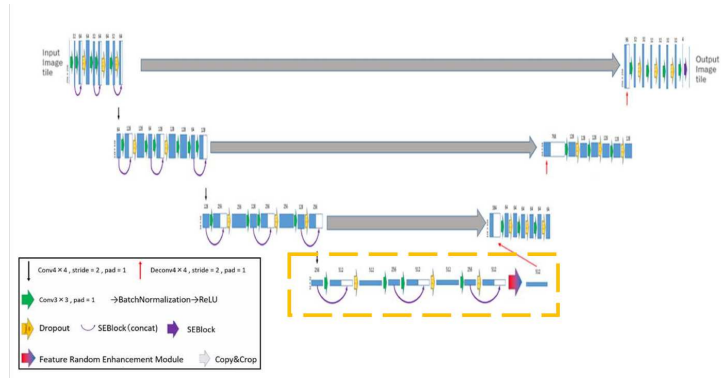}
\caption{U-Net with SE block}
\end{figure*}

\subsection{Implementation details}

To describe the proposed method for U-Net, the encoder’s output is enhanced by multiplying feature maps selected randomly at each epoch by X. The number of feature maps selected randomly is denoted as B. The feature maps are re-selected each epoch and the network weight is updated during training. 

The closest approaches is Dropout. Similarly, dropout is used only during training, and some neurons are randomly set to 0. If there is an element set to 0, the backpropagation stops at that element. It is a method to allow the ensembles. The proposed method differs from Dropout. We use an adjustable feature emphasis not setting to 0. This is to improve the case where there is a difference in the ease of updating between the near and far layers.

The proposed method can be implemented in addition to Dropout. However, this does not mean that Dropout will be replaced by the proposed method. Our method is difficult to implement in many layers because we determine adequate parameters X and B at each layer. Implementing it at the farthest layer from the loss function solves the problem presented in this paper and is the most effective.

Figure 3 shows the detailed description of the proposed method. In the proposed method, we multiply X by all values in the selected feature maps which are the end of encoder shown as yellow layer in Figure 1. This operation is performed only in training phase. The enhanced feature maps are selected randomly. Thus, all channels in encoder’s output are not enhanced. We need to select hyperparameters X and B appropriately. Hyperparameters were searched by using the optimization with Tree-structured Parzen Estimator (TPE) \cite{nineteen} which uses Bayesian optimization.

\section{\uppercase{Experiments}}

This section shows the experimental results of the proposed method. Section 4.1 and 4.2 describe the dataset and the network used in experiments. Experimental results are shown in section 4.3. In section 4.4, additional experiments are conducted for considerations.

\subsection{Dataset}

In this paper, we conduct experiments on two kinds of cell image datasets. The first dataset includes only 50 fluorescent images of the liver of a transgenic mouse expressing a fluorescent marker on the cell membrane and nucleus \cite{twenty}. The size of the image is 256 $ \ times $ 256 pixels and consists of three classes; cell membrane, cell nucleus, and background. We use 35 images for training, 5 images for validation, and 10 images for test. 

The second dataset includes 20 Drosophila feather images \cite{twentyone}. The size of the image is 1024 $ \ times $ 1024 pixels and consists of four classes; cell membrane, mitochondria, synapse, and background. Training and inference were performed by cropping 16 images of 256 × 256 pixels from one image without overlap due to GPU memory size. Intersection over Union (IoU) and Mean IoU were used as evaluation measure for both datasets. 

\subsection{Network}

The proposed method introduces a module that randomly enhances the features at the final layer of encoder during only training. We call it “Feature Random Enhancement Module”. Fig. 4 shows the U-Net used in this paper. As shown in Fig. 4, the proposed module was implemented on a standard U-Net with SE block. 
Some feature maps are selected from 512 feature maps at the farthest layer from the loss function  which is shown as the bottom right in Fig. 4 at training phase, and the value in the feature map is multiplied by X. 

\subsection{Results}

In all experiments, we trained all methods till 2000 epochs in which the learning converges sufficiently, and evaluation is done when the highest mIoU accuracy is obtained for validation images. We used softmax cross entropy. The hyperparameters B and X were searched 50 times using the Tree-structured Parzen Estimator algorithm (TPE) which seem to be a sufficient number. 

For comparison, U-Net with only SE block is evaluated. This is the baseline. U-Net with SE block without yellow layer in Fig.1 (the yellow dotted square in Fig. 4) is evaluated to present the problem that deep layers are difficult to train. The problem is that the model without those deep layers achieved higher accuracy than that with those deep layers. 

We also evaluate the U-net with SE block which uses SuperVision instead of the proposed module in order to show the effectiveness of Feature Enhancement module. SuperVision uses 1x1 convolution to change the number of channel to the number of classes at the end of encoder, and resize it to the size of input image and softmax cross entropy loss is computed. 
When we use SuperVision, we must optimize two losses; the first loss is standard softmax cross entropy loss at the final layer and the second loss is for supervision. In general, the balancing weight for two losses should be optimized.
\begin{center}
Loss=(1- $\lambda$)*Loss.1+$ \lambda $*Loss.2   (1) 
\end{center}
where is $ \lambda $ the balancing weight. The parameter is also optimized by TPE. The search was performed 15 times to find the appropriate parameter. Since $ \lambda $ is a single parameter, the number of searches is smaller than that of the two parameters B and X in our method.

First, we show the experimental results on the mouse cell dataset. Table 1 shows the results when we use B = 162 and X = 632 which gives the highest mean IoU for validation set. Table 1 shows that the accuracy of our method is improved in all classes in comparison with the conventional models. We confirmed 2.12\% improvement from the baseline. In addition, the accuracy of the proposed method is better than U-Net + SENet without deep layers. This result shows our method can train deep layers effectively.
In addition, the accuracy of U-Net + SENet without deep layers is better than U-Net + SENet. It shows that the deep layers of the U-Net + SENet used in the experiments is far from the loss and not updated well. 
The accuracy of mean IoU is not improved by U-Net with SuperVision ($ \lambda $ = 0.3257 determined by TPE) even if loss is computed at the end of encoder that our module is used. 
When we use dropout with the same percentage as the proposed method (Dropout rate = B/the number of filters = 162/512), it did not improve the accuracy.
\begin{table*}
\begin{center}
\caption{IoU of Transgenic mouse cell dataset}
\label{table:headings}
\begin{tabular}{c|ccc|c}
\hline
\ & menbrane[\%] & nuclear[\%] & background[\%] & mIoU[\%]\\
\hline
\hline
 {\scriptsize U-Net + SEblock}&37.78&65.75&74.96&59.50\\
 {\scriptsize U-Net + SEblock without deep layers}&39.15&66.84&75.11&60.36\\
 {\scriptsize U-Net + SEblock + Supervision}&39.23&64.99&73.34&59.19\\
 {\scriptsize Dropout (Same percentage as the proposed method)} &36.44&65.50&75.09&59.01 \\
 {\scriptsize Proposed method}&\textcolor{red}{40.53}&\textcolor{red}{67.58}&\textcolor{red}{76.75}&\textcolor{red}{61.62}\\
\hline
\end{tabular}
\end{center}
\end{table*}

\begin{table*}
\begin{center}
\caption{IoU of the Drosophila dataset}
\label{table:headings}
\begin{tabular}{c|cccc|c}
\hline
\ {\scriptsize } &  {\scriptsize menbrane[\%]} &  {\scriptsize nuclear[\%]} &  {\scriptsize background[\%]} &  {\scriptsize Synapus[\%]} &  {\scriptsize mIoU[\%]}\\
\hline
\hline
 {\scriptsize U-Net + SEblock}& {\scriptsize 91.80}& {\scriptsize 76.87}& {\scriptsize 76.89}& {\scriptsize 50.46}& {\scriptsize 73.98}\\
  {\scriptsize U-Net + SEblock without deep layers}& {\scriptsize 92.32}& {\scriptsize 78.24}& {\scriptsize 76.93}& {\scriptsize 51.31}& {\scriptsize 74.70}\\
 {\scriptsize U-Net + SEblock + Supervision}& {\scriptsize 92.39}& {\scriptsize 77.77}& {\scriptsize 78.24}& {\scriptsize 52.21}& {\scriptsize 75.15}\\
  {\scriptsize Dropout (Same percentage as the proposed method)}& {\scriptsize 92.40}& {\scriptsize 77.98}& {\scriptsize 78.70}& {\scriptsize 50.23}& {\scriptsize 74.83}\\
 {\scriptsize Proposed method}& {\scriptsize \textcolor{red}{92.93}}& {\scriptsize \textcolor{red}{78.71}}& {\scriptsize \textcolor{red}{78.02}}& {\scriptsize \textcolor{red}{58.14}}& {\scriptsize \textcolor{red}{76.95}}\\
\hline
\end{tabular}
\end{center}
\end{table*}

\begin{figure}
\centering
\includegraphics[height=4.0cm]{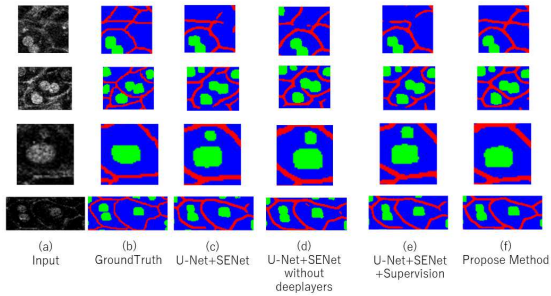}
\caption{Segmentation results on transgenic mouse cell images}
\end{figure}

\begin{figure}
\centering
\includegraphics[height=4.0cm]{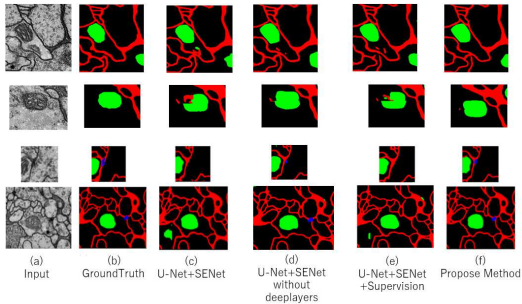}
\caption{Segmentation results on the Drosophila dataset}
\end{figure}

\begin{figure}
\centering
\includegraphics[height=4.0cm]{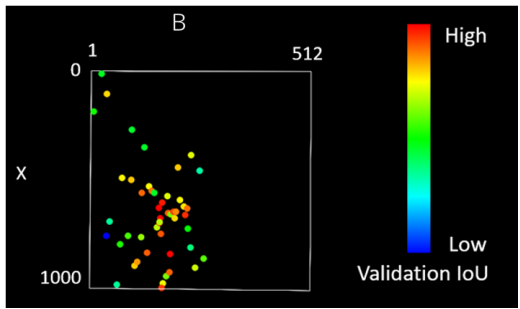}
\caption{TPE algorithm of Transgenic mouse cell}
\end{figure}

\begin{figure}
\centering
\includegraphics[height=4.0cm]{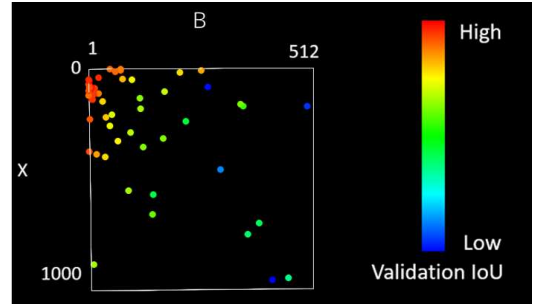}
\caption{TPE algorithm of Drosophila feather}
\end{figure}

Figure 5 shows the qualitative results. In Fig. 5, (a) is input image, (b) is ground truth, (c),(d) and (e) are the results by the conventional U-Net with SE block, U-Net with SE block without deep layers and U-Net with SE block + SuperVision, respectively, and (f) is the result by the proposed method. We see that cell image is blurred and it is difficult for not experts to assign class labels. This is because cells are killed by strong light and images are captured with low illuminace. 

In conventional method (d), there are many non- and over-detected cell membrane or nucleus. In addition, in conventional method (e), there are many undetected membranes. 
However, in the proposed method (f), more accurate segmentation results are obtained. This is because the proposed module enables to extract features from areas where training has not been done successfully in conventional methods. In addition, the method using SuperVision gave lower accuracy than the proposed method. We consider that the loss from the middle layer does not always give an intermediate representation for good segmentation result. 

\begin{table*}
\begin{center}
\caption{For additional experiment,IoU of Transgenic mouse cell dataset}
\label{table:headings}
\begin{tabular}{c|ccc|c}
\hline
\ & menbrane[\%] & nuclear[\%] & background[\%] & mIoU[\%]\\
\hline
\hline
{\scriptsize U-Net + SEblock}&37.78&65.75&74.96&59.50\\
{\scriptsize Proposed method(additional experiment)}&\textcolor{red}{40.43}&\textcolor{red}{67.34}&\textcolor{red}{73.71}&\textcolor{red}{60.49}\\
\hline
\end{tabular}
\end{center}
\end{table*}

Next, we show the experimental results on the Drosophila dataset. Table 2 shows the results when we use B = 8 and X = 250 when the highest mIoU is obtained for validation set. Table 2 shows that the accuracy of the proposed method is higher than that of the U-Net with SE block, and the mean IoU was improved 2.97\% by baseline. Furthermore, the accuracy was improved in comparison with the conventional models. In addition, the accuracy of U-Net + SENet without deep layers is better than U-Net + SENet. This result is the same as mouse cell data set. It reinforces the theory that the deep layers of the U-Net + SENet used in the experiments are not updated well. The results by U-Net with SuperVision ($ \lambda $ = 0.2781 determined by TPE) and Dropout with the same percentage as the proposed method (Dropout rate = B/the number of filters = 8/512) reinforce the theory.

Figure 6 shows qualitative results. In Fig. 6, (a) is the input image, (b) is ground truth, (c),(d) and (e) are the results by the U-Net with SE block, U-Net with SE block without deep layers and U-Net with SE block and SuperVision, and (f) is the results by the proposed method. In the Drosophila dataset, the image seems to contain enough information but it is difficult for ordinary people to assign correct class labels to each pixel. However, we confirmed that the proposed method (f) performs better segmentation for cell membrane and nucleus.

Figure 7 and 8 show the results of hyperparameter search using the TPE algorithm. The vertical and horizontal axes show the hyperparameters B and X in the proposed module. Red points show high mean IoU for validation set, and the blue points show low accuracy. We see that the TPE algorithm focuses on searching for places with high accuracy. Of course, optimal B and X depend on the dataset. However, we can find good hyperparameters by TPE. 

\begin{figure}
\centering
\includegraphics[height=5.0cm]{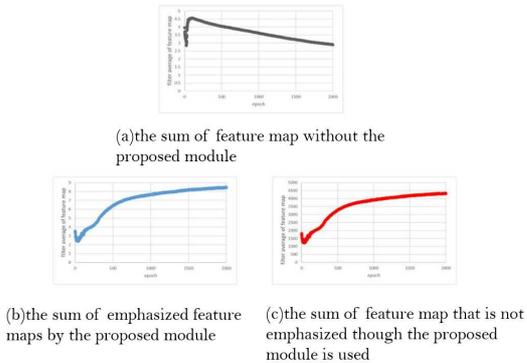}
\caption{Sum of feature map with/without Feature Enhancement module}
\end{figure}

\subsection{Additional Experiments}

The proposed method emphasizes some feature maps randomly at each epoch to prevent over-fitting. However, when we apply 10,000 times enhancement to the fixed ten filters during training, IoU accuracy was improved by about 1\% as shown in Table 3. This shows that the effect can be seen even if the emphasis is not performed randomly. Thus, we observed that the sum of the values in the feature map that ReLU function is adopted after convolution because we want to confirm the behavior of the enhanced and unenhanced filters. 

Figure 9 (a) shows the sum of feature map without the proposed module, (b) shows the sum of emphasized feature map by the proposed module. In (c), we used the proposed module but we computed the sum of non-emphasized feature map for the comparison with (b). In (a), the sum of feature map gradually decreases and the feature maps at the end of encoder are not used effectively. On the other hand, in (b) and (c), the sum of feature map increased through training. This means that the feature maps at the end of encoder have large value automatically and those features are used to obtain segmentation results. The proposed method emphasizes some feature maps randomly at each epoch to prevent over-fitting. Therefore, from the change of the value of (c) compared with (a), we see that the proposed module also has an effect on the feature maps that are not emphasized.

\section{\uppercase{Conclusion}}

In this paper, we introduced the Feature Random Enhancement Module which is enhanced feature map randomly during only training. We succeeded in improving the accuracy on cell image segmentation. We could propose the method for improving accuracy though the amount of computation during inference does not change.

A future work is to establish a method for deriving the parameters of the proposed module. Although TPE seems to be effective for parameter search from the results, it requires training for each parameter until the accuracy converges. Therefore, the computational cost for inference is fast but training takes longer time. Thus, we would like to study whether those parameters can be determined faster without convergence.

\section*{\uppercase{Acknowledgement}}
This work is partially supported by MEXT/JSPS KAKENHI Grant Number 18K111382 and 20H05427.

%
%

\bibliographystyle{apalike}
\end{document}